\documentclass[prl,twocolumn,final]{revtex4}

\usepackage{graphicx}

\begin{document}
\bibliographystyle{apsprl}


\title{The Antarctic climate anomaly and galactic cosmic rays}
\author{Henrik Svensmark}
\affiliation{Center for Sun Climate Research, Danish National Space
Center, Juliane Marie Vej 30, 2100 Copenhagen {\O}, Denmark}
\received{}

\begin{abstract}
It has been proposed that galactic cosmic rays may influence the
Earth's climate by affecting cloud formation. If changes in
cloudiness play a part in climate change, their effect changes sign
in Antarctica. Satellite data from the Earth Radiation Budget
Experiment (ERBE) are here used to calculate the changes in surface
temperatures at all latitudes, due to small percentage changes in
cloudiness. The results match the observed contrasts in temperature
changes, globally and in Antarctica. Evidently clouds do not just
respond passively to climate changes but take an active part in the
forcing, in accordance with changes in the solar magnetic field that
vary the cosmic-ray flux.
\end{abstract}

\maketitle Evidence has accumulated in recent years that the influx
of galactic cosmic rays, as modulated by solar magnetic activity,
influences the Earth's temperature by varying the cloudiness at low
altitudes\cite{Hsv:97,Hsv:98,Hsv:00}. Electrons liberated by muons
help to make the cloud condensation nuclei on which water droplets
form\cite{Svensmark:05}. There is now no reason to doubt that the
Earth's atmosphere acts like a natural cloud chamber that registers
the passing muons. What remains to be demonstrated is that the
resulting clouds affect the climate, and that is the purpose of this
paper.

Contradictory trends in temperature in Antarctica and the rest of
the world, which are evident on timescales from millennia to
decades, provide a strong clue to what drives climate change. The
southern continent is distinguished by its isolation and by its
unusual response to changes in cloud cover. While the rest of the
global surface is (on balance) cooled by clouds, they have a warming
effect on high-albedo
snowfields\cite{Ambach:74,Nakumura:88,Konzelmann:94,Bintanja:96,Nardino:00,Pavolonis:03}.
NASA's Earth Radiation Budget Experiment (ERBE)
\cite{Ramanathan:89,Barkstrom:90albedo} provided valuable data on
the effects of clouds at different latitudes. They can be
interpreted to show that, if changes in cloudiness drive climate
change, the anomalous behavior of Antarctica is predictable

\begin{figure}[h]
\includegraphics[scale=0.5]{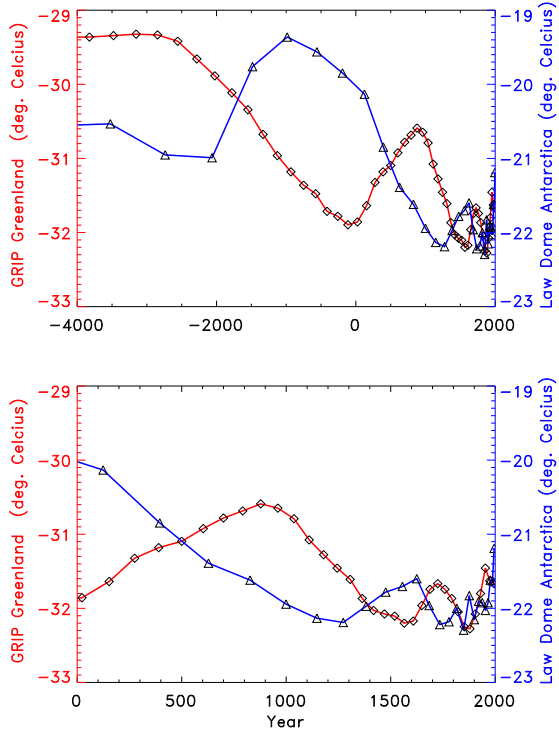}
\caption{ Ice temperatures from the GRIP site in Greenland
(73$^{\circ}$N, 38$^{\circ}$W)(red) and Law Dome in Antarctica
(67$^{\circ}$N, 112$^{\circ}$E)(blue) using borehole thermometry
data from Dahl-Jensen et al.\cite{Dahl-Jensen:98,Dahl-Jensen:99}.
The Antarctic climate anomaly is particularly conspicuous during the
cold period of the first millennium BC and the warm Viking Age c.
1000 AD. The apparent increase in frequency of the oscillations is
not real but is due to a smoothing of the older temperature records
by thermal conduction in the ice.} \label{fig:1}
\end{figure}

Borehole temperatures in the ice sheets spanning the past 6000 years
show Antarctica repeatedly warming when Greenland cooled, and vice
versa (Fig. \ref{fig:1}) \cite{Dahl-Jensen:98, Dahl-Jensen:99}.
North-south oscillations of greater amplitude associated with
Dansgaard-Oeschger events are evident in oxygen-isotope data from
the Wurm-Wisconsin glaciation\cite{Blunier:01}. The phenomenon has
been called the polar see-saw\cite{Blunier:01,Shackleton:01}, but
that implies a north-south symmetry that is absent. Greenland is
better coupled to global temperatures than Antarctica is, and the
fulcrum of the temperature swings is near the Antarctic Circle. A
more apt term for the effect is the Antarctic climate anomaly.

Attempts to account for it have included the hypothesis of a
south-flowing warm ocean current crossing the
Equator\cite{Knutti:04} with a built-in time lag supposedly intended
to match paleoclimatic data. That there is no significant delay in
the Antarctic climate anomaly is already apparent at the
high-frequency end of Fig.\ (\ref{fig:1}). While mechanisms
involving ocean currents might help to intensify or reverse the
effects of climate changes, they are too slow to explain the almost
instantaneous operation of the Antarctic climate anomaly.

The contrasts in temperature trends are not predicted by variations
in the concentrations of carbon dioxide and other greenhouse gases.
These gases diffuse throughout the atmosphere as far as the South
Pole. When they increase, climate models predict simultaneous and
strong warming at both ends of the Earth. There is no direct
physical reason why this forcing should operate differently in the
two polar regions.

\begin{figure}[h]
\vspace{0cm}
\includegraphics[scale=0.5]{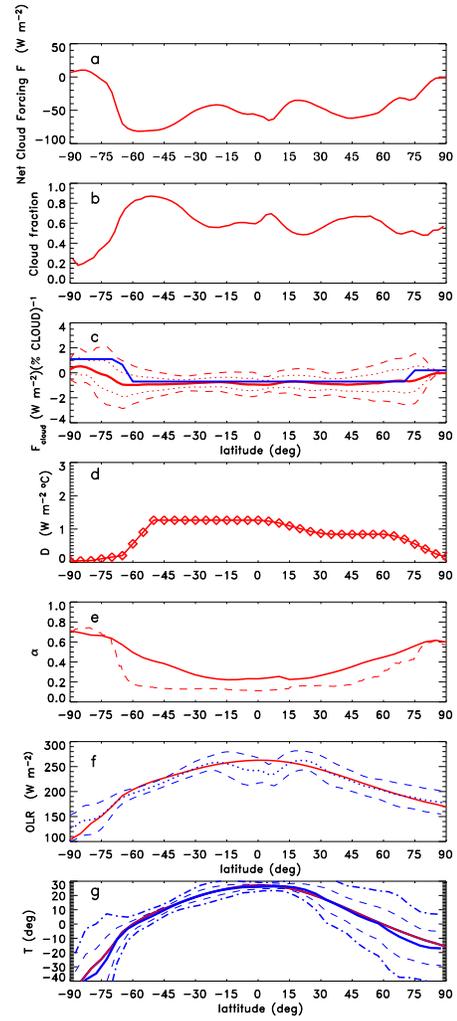}
\caption{ Satellite observations (ERBE data [ref.]) relevant to
calculations of the forcing effect of a change in cloudiness at
different latitudes. a: Net cloud forcing by latitude. b: Cloud
fraction by latitude. c: Net cloud forcing from a 1 \% uniform
change in cloud cover (red curve), together with one and two sigma
variations (broken lines). The blue curve is the $F_{\rm{cloud}}(q)$
function used in the model given by Eq.\ (\ref{EBM}).
Parametrization used in the model given by Eq.\ (\ref{EBM}) as a
function of latitude. d: Eddy-diffusion constant. e: Albedo solid
red line. Broken red line is clear sky albedo. Finally the model
Eq.\ (\ref{EBM})results compared with observations f: Out going
long-wave radiation. Blue line observation based on ERBE with root
mean square variances (dotted blue lines), red line model result. g:
Observed surface temperatures solid blue with one and two sigma
variations. Red curve model result.} \label{fig2}
\end{figure}

The simplest and most immediate explanation of the Antarctic climate
anomaly comes from cloud forcing. In most climate models, clouds are
passive participants responding to changes due to other forcing
agents. If, on the other hand, changes in cloudiness and GCR's drive
the Earth's climate\cite{Hsv:97,Hsv:98,Hsv:00} the Antarctic climate
anomaly is the exception that proves the rule.

Clouds warm the underlying surface by trapping the outgoing
long-wave radiation, and cool it by reflecting the short-wave
radiation from the Sun. In general the cooling effect is greater
than the heating effect, resulting in a net cooling of the Earth of
the order of 15 W/m$^2$. A small percentage change in cloud cover
can therefore result in significant forcing.

\begin{figure}[h]
\includegraphics[scale=0.5]{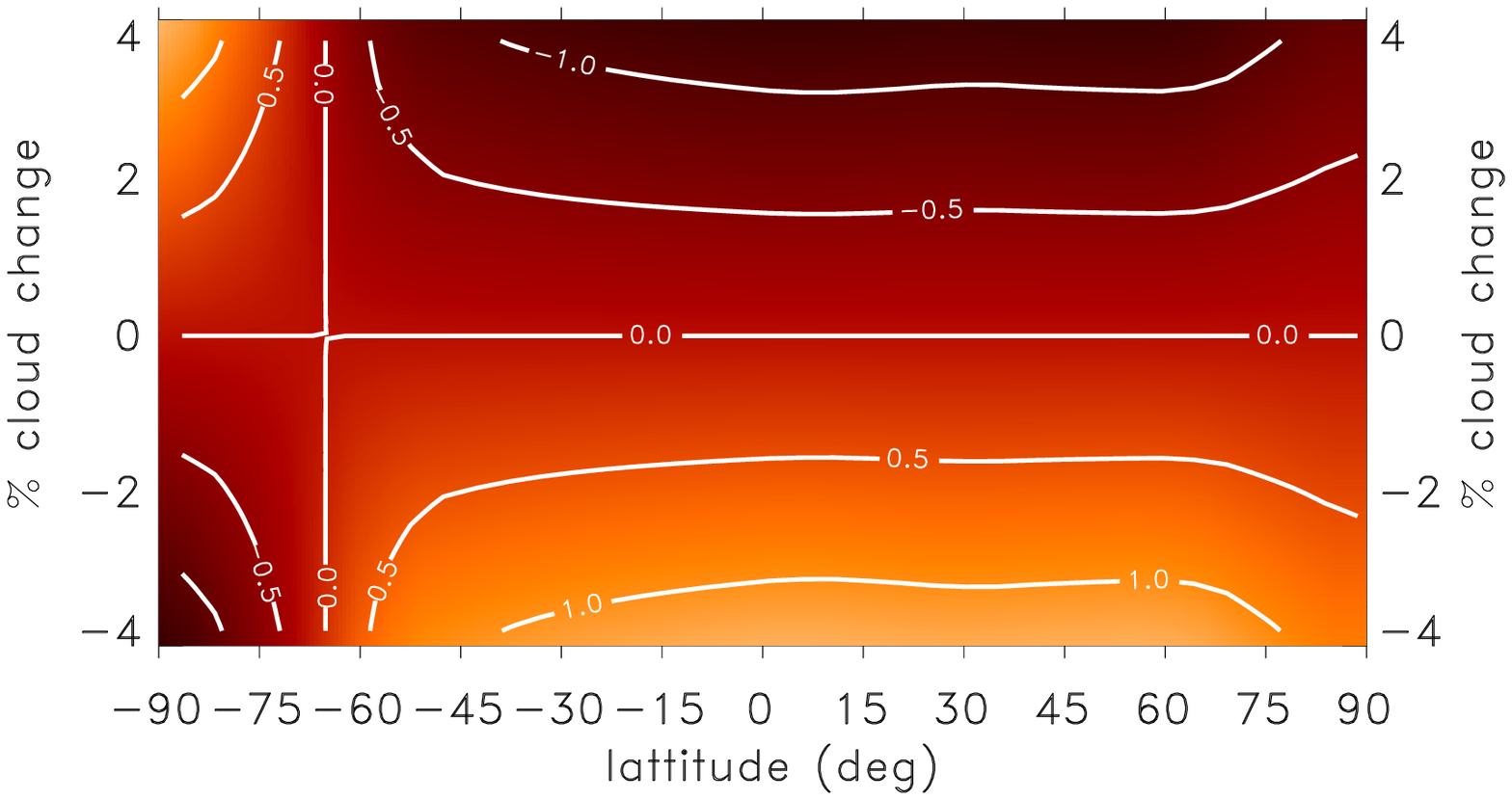}
\caption{Top panel: Calculated responses of surface temperatures at
different latitudes to uniform percentage changes in cloud cover
($\Delta_0$) from $-$4 \% to +4 \%. The contour lines show
temperature changes in fractions of a degree Celsius. Note the
saddle-point of the Antarctic climate anomaly at latitude $-$63.}
\label{fig:3}
\end{figure}

The cooling effect is not evenly distributed. As shown in Fig.\-
(\ref{fig2} a) it is minimal around the Equator and increases
towards the mid-latitudes. In polar regions the clouds can have a
warming effect if their re-radiation of long-wave energy downwards
dominates over the loss of short-wave solar energy blocked by the
clouds. This warming has been well recorded on the surface in both
the Arctic and
Antarctic\cite{Ambach:74,Nakumura:88,Konzelmann:94,Bintanja:96,Nardino:00,Pavolonis:03}
\begin{figure}[h]
\includegraphics[scale=0.45]{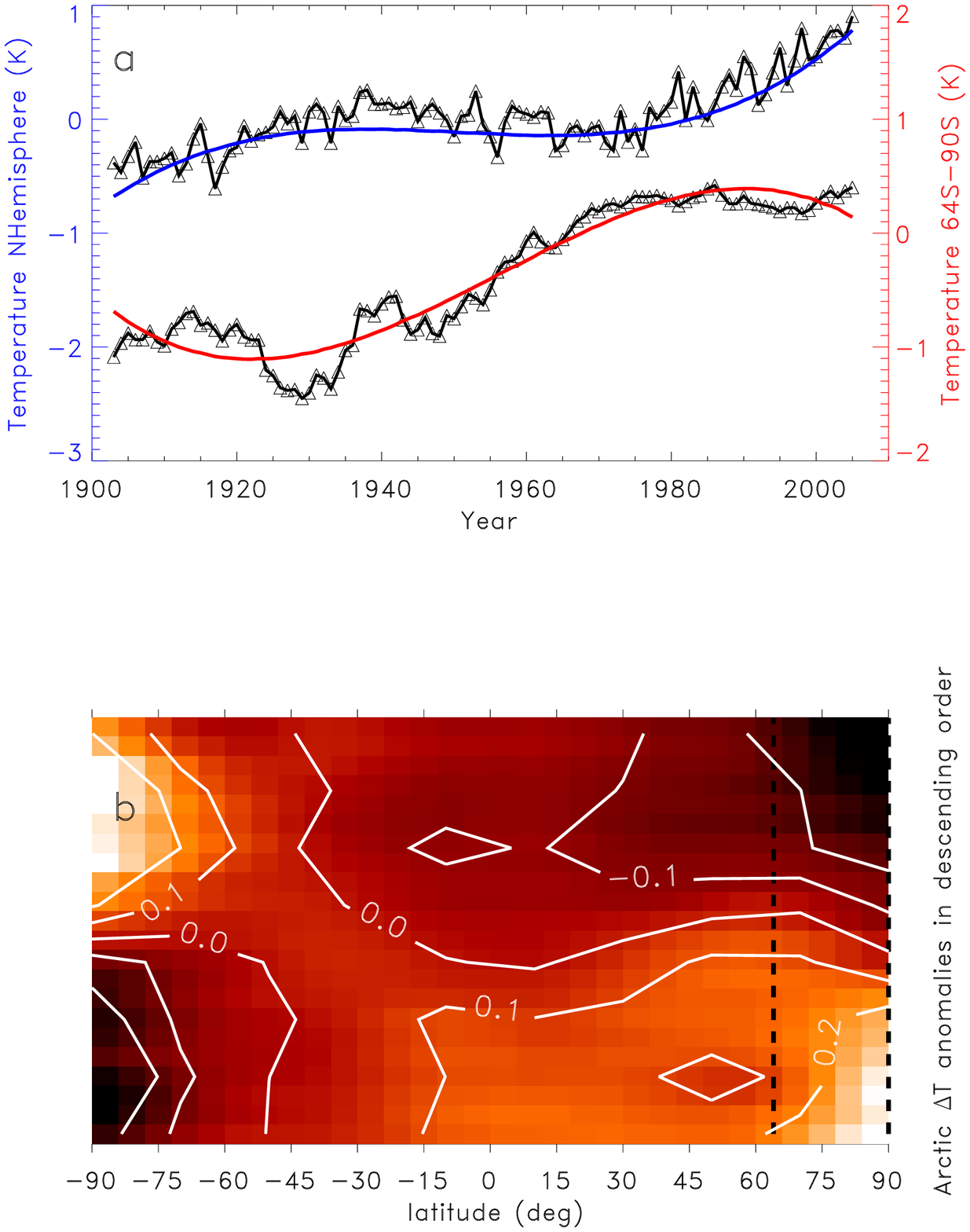}
\caption{{\bf a}: The Antarctic climate anomaly during the past 100
years is apparent in this comparison of the annual surface
temperature anomalies for the northern hemisphere and Antarctica
(64S-90S), from the NASA-GISS compilations. The Antarctic data has
been averaged over 12 years to minimize the temperature
fluctuations. The blue and red line are fourth order polynomial fit
to the northern hemisphere data and the Antarctica (64S-90S) data,
respectively. If the two curves were not offset by 1 $^\circ$C, for
clarity, the polynomial trendlines would cross and re-cross around
1910, 1955 and 1995, in the same manner as the millennial-scale
curves of the Antarctic climate anomaly in Fig. \ref{fig:1}. In the
cloud-forcing picture, the overall warming of the Antarctic during
the 20th Century is probably due to negative feedback from the
greenhouse action of increasing water vapor reaching the polar
atmosphere. {\bf b}: Visualization of the Antarctic climate anomaly
using the NASA-GISS compilations of annual temperature anomalies
$\Delta T$ for 1903-2005 in 102 time intervals and 8 zonal bands
($\pm$(90--64) $\pm$(64--44) $\pm$(44--24) $\pm$(24--0)). First, the
linear trend in each zonal time series is removed. Next, the data
for the 102 time intervals in the Arctic series are arranged in
descending order of $\Delta T$ on the right side of the figure,
poleward of 64 deg, between the black dotted lines. Then the $\Delta
T$'s for the corresponding time intervals are plotted for each of
the other zones. Finally the surface is smoothed and contour lines
of $\Delta T$ in $^0$C are added. Note the N-S contrast in every
time interval, and the topological similarity to the computed cloud
forcing in Fig. (3).} \label{fig:4}
\end{figure}

Figure (\ref{fig2}a) also shows that the polar warming effect of
clouds is not symmetrical, being most pronounced beyond 75$^\circ$S.
In the Arctic it does no more than offset the cooling effect,
despite the fact that the Arctic is much cloudier than the Antarctic
(Fig.\ (\ref{fig2}b)). The main reason for the difference seems to
be the exceptionally high albedo of Antarctica in the absence of
clouds.

The ERBE satellite data shown in Fig.\ (\ref{fig2}) provide a basis
for computing the cooling and warming effects of small changes in
cloud cover. The changes are assumed to be the same at all
latitudes. In order to achieve a forcing that is independent of
cloud fraction, the net cloud forcing in Fig.\ (\ref{fig2}) (a) is
normalized with the cloud fraction in Fig.\ (\ref{fig2}) (b) to
derive the cloud forcing at different latitudes for a 1 \% increase
in cloud fraction, as shown in red in Fig.\ (\ref{fig2}) (c).
Finally, the blue curve in Fig. (\ref{fig2}) (c) is a simplified
version of the cloud forcing used as ($F_{\rm{cloud}} (q)$) in the
calculation that follows. A simple model balances incoming
short-wave energy from the sun and outgoing long-wave radiation, and
allows diffusion of heat between neighboring latitude bands. It
takes the form\cite{North:75}
\begin{eqnarray}
\label{EBM} c \;\partial_t U(x,t) = \nonumber
\partial_x (1-x^2) D(x) \partial_x U(x,t)  \\
+ Q S(x)(1- \alpha(x))- A - B U(x,t) - \Delta_0 F_{\rm{cloud}} (x)
\end{eqnarray}
where $x = \sin q$ and q is latitude. $U(x,t)$ is the temperature at
latitude $x = \sin q$, $t$ is time, $c$ is the heat capacitance,
$D(x)$ is an eddy diffusion constant that varies with latitude and
is shown in Fig.\ (2d), $Q$ is the solar constant (1370 W/m$^2$),
$S(x)$ is the average fraction of solar radiation at latitude $x$,
and $\alpha (x)$ is the albedo at latitude $x$ shown in Fig.\ (2e)
solid red curve\cite{Barkstrom:90albedo}. $\Delta_0$ is the
percentage change in cloud cover, compounded with $F_{\rm{cloud}}
(q)$ from Fig. (\ref{fig2}) (c) The adjustable parameters $A$ and
$B$ determine the long-wave loss at latitude $q$. The resulting
values are $Q$= 1370 W/m$^2$, $A$ = 215 W/m$^2$, and $B$ = 1.7
W/m$^2$$^\circ$C$^{-1}$. Since Eq.\ (\ref{EBM}) is solved for steady
state the value of $c$ is not important and was set to $1$
numerically. The procedure used, was to adjust the eddy diffusion
constant so both the model outgoing longwave
radiation\cite{Ramanathan:89,Barkstrom:90albedo} and the model
surface temperature, as a function of latitude fitted
observations\cite{Jones:99}. The results of this procedure are shown
in Fig.\ (2e) and Fig.\ (2g).

The calculated temperature changes in response to a uniform
variation of the cloud cover are shown in Fig.\ (\ref{fig:3}). The
sign of the response (warming or cooling) reverses at a saddle-point
which in this simple model is around $-63$ deg. latitude. The
climate sensitivity of the model is approximately $\gamma =$ 0.5
W$^{-1}$m$^{2}$C$^{}$. For an $\pm$ 4 \% change in cloud cover, the
variation in temperature is about 2 $^\circ$C at $+80$ deg. and
$\approx 1.5$ C (opposite sign) at $-80$ deg. Such changes can
account for the temperature excursions of 2$^\circ$C or less between
Greenland and Antarctica during the past 6000 years, shown in Fig.
(\ref{fig:1}).

Figure (\ref{fig:3}) also predicts that a reduction in cloud cover
of about 8 \% is sufficient to warm most of the globe by almost
2$^\circ$C, which is in line with other estimates of cloud forcing
during the 20th Century\cite{Hsv:97}. The effect is seen in the
upper curve of Fig.\ (\ref{fig:4}a) (NASA-GISS data\cite{GISS:00}).
In a cloud interpretation the hesitations and advances in cloud
reduction since 1900 follow the well-known changes in solar
activity\cite{Hsv:97,Lockwood:99}. The lower curve in Fig.\
(\ref{fig:4}a) shows the corresponding changes in Antarctica, and
the operation of the Antarctic climate anomaly is plain to see. Note
especially the fall in Antarctic temperatures in the 1920s
contrasting with a surge in global temperatures, and the marked rise
1950--70 when global temperatures fell.

Unfortunately there do not exist a record of cloud cover that could
verify the connection between clouds and climate directly. However
satellite data of Earths cloud cover from the International
Satellite Cloud Climate Project (ISCCP) covering 1983 -- 2005 show
that the large scale temporal variations are distributed fairly
evenly over the globe\cite{Hsv:00}, and for example cloud variations
over oceans and clouds over Antarctica have similar temporal
evolution.

A remaining question about Fig.\ (\ref{fig:4}a) is how the Antarctic
temperatures were dragged upwards, to take part in the general
warming, despite the long-term reduction in cloud cover. A natural
mechanism must be in operation, because similar correspondences
(though in the context of an overall global cooling) are seen on the
millennial scale in the Greenland and Antarctic ice temperatures in
Fig.\ (\ref{fig:1}). The most likely explanation is the diffusion of
water vapor to the Antarctic atmosphere, as a result of the
increased capacity for water vapor in warmer global
atmosphere\cite{Kirk:99,Jain:05}. The strong natural greenhouse
effect of the additional water vapor would amplify the effect of
cloud forcing globally (positive feedback) and over-ride it in
Antarctica (negative feedback).

Finally, it is possible to map zonally averaged NASA-GISS
temperature data for the past century so that the Antarctic Climate
Anomaly become apparent. By removing the liner trend of the 8 zonal
temperature bands series covering the years 1903-2005, and
subsequently sort all the 8 temporal time series after the
descending temperatures of the 64-90 latitude band, Fig.\
(\ref{fig:4}b) is obtained. This figure is very similar to Fig.\
(\ref{fig:3}), with a saddle point structure in the souther
hemisphere. In this case the node of the saddle point seems to be at
somewhat lower latitudes. This effect could real and caused by
clouds over the ocean. A resent study found that poleward of
$-$58.75 (observational limit) had a heating effect on the surface
over most of the year\cite{Pavolonis:03}.

Cloud forcing is by far the most economical hypothesis that explains
the patterns of Fig.\ (\ref{fig:1}) and Fig.\ (\ref{fig:4}), as well
as the bigger see-saw effects in Wurm-Wisconsin
times\cite{Blunier:01}. There is plainly scope for more detailed
modeling of the Antarctic climate anomaly on various timescales.

Meanwhile, a chain of evidence appears to be complete, which links
low-level clouds to the well-known modulation of galactic cosmic-ray
intensity by solar magnetic activity, to the detected influence of
galactic cosmic rays on cloudiness\cite{Hsv:97,Hsv:98,Hsv:00}, and
also to experimental evidence that electrons set free by passing
muons help to make aerosols the pre-curser to cloud condensation
nuclei at low altitudes\cite{Svensmark:05}. The roles of cosmic rays
and clouds as active players in climate change therefore merit
closer attention in general climate modeling and in solar and
heliospheric physics, with special regard to the high-energy
galactic cosmic rays that ionize the lower atmosphere. Physics
history comes full circle. More than 100 years ago, C.T.R. Wilson
developed the cloud chamber to try to understand natural clouds but
he was diverted by his detection of ionizing particles. In 1937 the
first known muons turned up in a cloud chamber\cite{Street:37}. Now
Wilson's initial purpose is fulfilled in a fresh understanding of
the physics of natural low-level clouds, and in the evidence
presented here for its relevance in the real world.


\begin{acknowledgments}

\end{acknowledgments}

\end{document}